\documentstyle[11pt,newpasp,twoside,epsf]{article}
\def\dexkpc{~dex~kpc$^{-1}${}}

\def\edcomment#1{\iffalse\marginpar{\raggedright\sl#1\/}\else\relax\fi}
\reversemarginpar

\begin{document}
\title{Galactic metallicity distribution from open clusters}
\author{L. Chen, J.L. Hou \& J.J. Wang}
\affil{Shanghai Astronomical Observatory, CAS, Shanghai
200030,China}

\begin{abstract}
We have compiled two new Galactic open cluster catalogues. The
first one has 119 objects with age, distance and metallicity data,
while the second one has 144 objects with both absolute proper
motion and radial velocity data, of which 45 clusters also with
metallicity data available. An iron radial gradient of about
$-$0.063$\pm$0.008 \dexkpc \ is obtained from our sample. By
dividing clusters into age groups, we show that iron gradient was
steeper in the past. A disk age-metallicity relation could very
probably exist based on the current sample.

\end{abstract}

\section{Introduction}

Open clusters(OCs) have long been used to trace the structure and
evolution of the Galactic disk. Since (OCs) could be relatively
accurately dated and can be seen from large distance, their [Fe/H]
values serve an excellent tracer to the abundance gradient along
the Galactic disk as well as many other important disk properties,
such as abundance gradient evolution, disk age and so on.

In this paper, we give some statistical results of the open
cluster based on our up-to-date sample, mainly relating to the
Galactic abundance gradient and age-metallicity relation. Details
can be found in Chen, Hou \& Wang (2003).

\section{The catalogues }

We have compiled two new catalogues of the Galactic open clusters.
The first one (CAT 1) lists 119 clusters parameters for distance,
age and metallicity. The age, distance and reddening information
and  most iron abundance data are from Dias et al. (2002). Thus
far CAT 1 provides a most complete open cluster sample concerning
the iron abundance, distance and age parameters together. This
sample could provide statistically more significant information
concerning the radial iron gradient as well as its evolution, etc.
In the second catalogue (CAT 2), we have listed observed
kinematical data from literature for 144 clusters, with both
radial velocity and mean proper motion available. The mean radial
velocity data are mostly from a compilation in WEBDA database. The
absolute proper motion of clusters, based on the Hipparcos system,
are from Baumgardt et al.(2000) and Dias et al.(2001). In fact,
the above observed kinematic information constitute a
sub-catalogue of that of Dias et al. (2002). But in CAT 2 we have
further calculated the three dimensional velocity of open clusters
by combing with radial velocity and mean absolute proper motion
data. In addition, for each cluster, age and iron abundance data
are also listed whenever available. Two catalogues are available
online.

\section{Some statistical results}

\subsection{The abundance gradients }

The first radial metallicity gradient using open clusters was
given by Janes (1979), with derived gradient of $-$0.05 \dexkpc.
Since then a lot of works have been done on the field.

The existence of radial iron abundance gradients is also confirmed
by our up-to-date sample. By least-square fitting, we derived an
radial abundance gradient of $-$0.063 $\pm$ 0.008 \dexkpc, which
agrees well with most of the previous open cluster results.

\subsection{Gradient evolution}

We have calculated gradients for two sub-samples with cluster age
$<$ 0.8 Gyr (80 clusters) and $\ge$ 0.8 Gyr (38 clusters),
respectively. The fitting results are $-$0.024 $\pm$ 0.012 \dexkpc
\ for younger clusters, $-$0.075 $\pm$ 0.013 \dexkpc \ for older
ones. If we take the mean age for the youngest and oldest clusters
as 0.00 Gyr and 6.00 Gyr in our sample, we can estimate an average
flattening rate of 0.008 \dexkpc Gyr$^{-1}$ during the past 6 Gyr.
This rate is quite consistent with that from recent PN data for
[O/H] (Maciel et al. 2002), and is also in good agreement with the
model result of Hou et al. (2000).

\subsection{Disk age-metallicity relation}

The age-metallicity relation(AMR) for the Galactic disk provides
useful clues about the chemical evolution history of the Milky
Way, and also put an important constraint on the theoretical
models of the disk.

Based on our new open cluster catalogue with much more objects, we
found that, statistically, the space distributions of open
clusters are very likely imply the existence of age-metallicity
relation in the Galactic disk. However, in order to have more
definite conclusion, observational efforts should be added in
finding more older clusters.

\acknowledgments{This research was supported by the NSFC
(No.19873014, No.10173017, No.10133020) and NKBRSFG 19990754.

\end{document}